\documentclass[a4paper]{article}

\pdfoutput=1
\usepackage[utf8]{inputenc}
\usepackage[super,sort&compress,comma, square]{natbib}
\usepackage{tikz}
\usepackage{tablefootnote}
\usepackage{gensymb}
\usepackage[english]{babel}
\usepackage[T1]{fontenc}
\usepackage{csquotes}
\usepackage{doi}
\usepackage[a4paper,top=3cm,bottom=2cm,left=3cm,right=3cm,marginparwidth=1.75cm]{geometry}
\usepackage{titling}
\usepackage{parskip}

\usepackage[version=4]{mhchem}
\usepackage{amsmath}
\usepackage{graphicx}
\usepackage{hyperref}
\hypersetup{colorlinks=true, allcolors=blue}
\usepackage{authblk}
\usepackage{multirow}
\usepackage{nicematrix,enumitem,booktabs}
\usepackage{nomencl}
\usepackage{booktabs}
\usepackage{tabularx}
\usepackage{textcomp}
\usepackage{amsfonts}
\usepackage{outlines} %%same
\usepackage{subfig}
\usepackage[version=4]{mhchem}
\usepackage{eurosym}
\usepackage[locale = US]{siunitx}
\usepackage{bm}
\usepackage[normalem]{ulem} %%temporary, for organization purposes
\usepackage{pythonhighlight}
\usepackage{graphicx}
\usepackage{adjustbox}
\usepackage{tablefootnote}

\usepackage{pifont}% http://ctan.org/pkg/pifont
\newcommand{\cmark}[1][]{\textcolor{green!80!black}{#1\quad\ding{52}}}
\newcommand{\xmark}[1][]{\textcolor{red}{#1\quad\ding{53}}}
\newcommand{\partialcheck}[1]{\textcolor{orange}{\quad\ding{52}(#1\%)}}
\date{}

\newcommand{\numpatterns}{92,552 }
\newcommand{\numlabeled}{2179 }

\DeclareSIUnit{\myeuro}{\text{\euro}}
\DeclareSIUnit\angstrom{\text {Å}}
\usepackage[final]{changes}

\title{opXRD: Open Experimental Powder X-ray Diffraction Database}

\newcommand{\ITI}{1}
\newcommand{\INT}{2}
\newcommand{\HKUSTGZ}{3}
\newcommand{\USC}{4}
\newcommand{\LBLMF}{5}
\newcommand{\LBLALS}{6}
\newcommand{\Hoffmann}{7}
\newcommand{\LBLCS}{8}
\newcommand{\KFUPMMSE}{9}
\newcommand{\KFUPMIRCIMR}{10}
\newcommand{\CPT}{11}
\newcommand{\EMPA}{12}
\newcommand{\FZJIEMD}{13}
\newcommand{\KITICRT}{14}

\author[\ITI,\INT]{Daniel Hollarek}
\author[\ITI,\INT]{Henrik Schopmans}
\author[\ITI,\INT]{Jona Östreicher}
\author[\ITI,\INT]{Jonas Teufel}
\author[\HKUSTGZ]{Bin Cao}
\author[\USC]{Adie Alwen}
\author[\INT]{Simon Schweidler}
\author[\LBLMF]{Mriganka Singh}
\author[\LBLMF,\LBLALS]{Tim Kodalle}
\author[\Hoffmann]{Hanlin Hu}
\author[\LBLCS]{Gregoire Heymans}
\author[\KFUPMMSE,\KFUPMIRCIMR]{Maged Abdelsamie}
\author[\CPT]{Arthur Hardiagon}
\author[\EMPA]{Alexander Wieczorek}
\author[\EMPA]{Siarhei Zhuk}
\author[\FZJIEMD]{Ruth Schwaiger}
\author[\EMPA]{Sebastian Siol}
\author[\CPT]{François-Xavier Coudert}
\author[\KITICRT]{Moritz Wolf}
\author[\LBLMF]{Carolin M. Sutter-Fella}
\author[\INT]{Ben Breitung}
\author[\USC]{Andrea M. Hodge}
\author[\HKUSTGZ]{Tong-yi Zhang}
\author[\ITI,\INT,*]{Pascal Friederich}

\affil[\ITI]{Institute of Theoretical Informatics, Karlsruhe Institute of Technology (KIT), 76131 Karlsruhe, Germany.  E-mail: pascal.friederich@kit.edu} % Informatik-Hauptgebäude, Am Fasanengarten 5, 76131 Karlsruhe
\affil[\INT]{Institute of Nanotechnology, Karlsruhe Institute of Technology (KIT), 76131 Karlsruhe, Germany} % Hermann-von-Helmholtz-Platz 1
\affil[\HKUSTGZ]{Guangzhou Municipal Key Laboratory of Materials Informatics, Advanced Materials Thrust, Hong Kong University of Science and Technology (Guangzhou) (HKUST), Guangzhou 511400, China} % 1 Du Xue Road,
\affil[\USC]{Department of Chemical Engineering and Materials Science, University of Southern California (USC), Los Angeles CA 90089, USA} % 925 Bloom Walk
\affil[\LBLMF]{Molecular Foundry Division, Lawrence Berkeley National Laboratory (LBNL), Berkeley 94720 CA, USA}  %67 Cyclotron Rd, Berkeley, CA 94720, United States
\affil[\LBLALS]{Advanced Light Source, Lawrence Berkeley National Laboratory,  Berkeley 94720 CA, USA}  % 6 Cyclotron Rd, Berkeley, CA 94720, United States
\affil[\Hoffmann]{Hoffmann Institute of Advanced Materials, Shenzhen Polytechnic, Shenzhen 518055, China} % Liuxian Street NO.7098, Shenzhen 518055, China
\affil[\LBLCS]{Lawrence Berkeley National Laboratory (LBNL), Chemical Sciences Division, Berkeley 94720 CA, USA}  % 1 Cyclotron Rd, Berkeley, CA 94720, United States
\affil[\KFUPMMSE]{Material Science and Engineering Department, King Fahd University of Petroleum and Minerals (KFUPM), Dhahran 31261, Saudi Arabia} % 846R+8C2, Academic Belt Rd, KFUPM, Dhahran 34463, Saudi Arabia
\affil[\KFUPMIRCIMR]{Interdisciplinary Research Center for Intelligent Manufacturing and Robotics, King Fahd University of Petroleum and Minerals (KFUPM), Dhahran 31261, Saudi Arabia} %Academic Belt Road, Dhahran 31261, Saudi Arabia
\affil[\CPT]{Chimie ParisTech, PSL University, CNRS, Institut de Recherche de Chimie Paris, 75005 Paris, France} % 11 Rue Pierre et Marie Curie, 75005 Paris, France
\affil[\EMPA]{Empa–Swiss Federal Laboratories for Materials Science and Technology (EMPA), 8600 Dübendorf, Switzerland}  % Ueberlandstrasse 129, 8600 Dübendorf, Switzerland
\affil[\FZJIEMD]{Institute of Energy Materials and Devices, Forschungszentrum Juelich GmbH, 52425 Juelich, Germany} %Wilhelm-Johnen-Straße, 52428 Jülich
\affil[\KITICRT]{Engler-Bunte-Institut \& Institute of Catalysis Research and Technology, Karlsruhe Institute of Technology (KIT), Karlsruhe, Germany}  %Hermann-von-Helmholtz-Platz 1
\affil[*]{Corresponding author: pascal.friederich@kit.edu}

\begin{document}
\maketitle

\begin{abstract}
Powder X-ray diffraction (pXRD) experiments are a cornerstone for materials structure characterization.
Despite their widespread application, analyzing pXRD diffractograms still presents a significant challenge to automation and a bottleneck in high-throughput discovery in self-driving labs.
Machine learning promises to resolve this bottleneck by enabling automated powder diffraction analysis.
A notable difficulty in applying machine learning to this domain is the lack of sufficiently sized experimental datasets, which has constrained researchers to train primarily on simulated data. However, models trained on simulated pXRD patterns showed limited generalization to experimental patterns, particularly for low-quality experimental patterns with high noise levels and elevated backgrounds.
With the Open Experimental Powder X-Ray Diffraction Database (opXRD), we provide an openly available and easily accessible dataset of labeled and unlabeled experimental powder diffractograms.
Labeled opXRD data can be used to evaluate the performance of models on experimental data and unlabeled opXRD data can help improve the performance of models on experimental data, e.g. through transfer learning methods.
We collected \numpatterns diffractograms, 2179 of them labeled, from a wide spectrum of materials classes.
We hope this ongoing effort can guide machine learning research toward fully automated analysis of pXRD data and thus enable future self-driving materials labs.
\end{abstract}

\newpage
\section{Introduction}\label{sec:Introduction}
The advent of high-throughput experiments holds the prospect of significantly accelerating the speed of materials discovery\cite{Liu2019}. The synthesis and characterization of novel materials are becoming increasingly efficient and automated, increasing the throughput of samples in experimentation pipelines\cite{MacLeod2019, Ludwig2019, Ozaki2020}.

After fabricating a new material, a number of analysis techniques can be used to characterize the sample. One method that can be used for phase identification, phase quantification, grain size characterization, and to determine the crystal structure of a new material is powder X-ray diffraction (pXRD). 
When using pXRD measurements, crystal structures are typically determined through Rietveld refinement. In Rietveld refinement, an initial crystal structure model is fitted to the observed diffractogram by iteratively updating the structural model. Each update of the structural model seeks to minimize the difference between the observed diffractogram and the diffractogram simulated from the current structural model \cite{Dinnebier2019, Cano2021}. As Rietveld refinement is a local optimization method, the result of the refinement procedure is generally only as good as the initial structural model the process started from.

Manually performing Rietveld refinement is time-consuming and often requires expert knowledge. It is not scalable to the degree required to keep up with advances in throughput and efficiency in other steps of the experimentation pipeline. The refinement process requires the operator to determine an initial structural model from which the refinement can start and as well as initial values for parameters that characterize the background \cite{mccusker1999}. The structural model is usually obtained using search-match software, which identifies crystal structures with similar powder diffraction patterns from a database of crystal structures with accompanying powder diffraction patterns. However, an initial structural model obtained from such a database is not guaranteed to lead to an accurate structure solution through Rietveld refinement, especially not for novel structures. Additionally, attempting to refine all crystal structure parameters at once is known to lead to unphysical results\cite{Ozaki2020}. Hence parameters are refined iteratively, with each iteration only refining a limited set of parameters. Finding the correct order in which to refine structure parameters and finding the correct values for initial background parameters both present problems that add to the difficulty of the refinement process.

Machine learning has the potential to speed up the manual analysis of powder diffractograms and keep pace with an automated high-throughput experimentation environment\cite{Agrawal2019, Surdu2023}.
Models can be either trained to predict crystal structure information directly given a diffractogram, or they can be used to automate the conventional refinement workflow. In the latter case a model would first predict an initial crystal structure \cite{Surdu2023} which is then refined by a second model trained to perform the refinement process \cite{Feng2019}. So far due to an absence of labeled datasets with experimental diffractograms\cite{Wang2020}, machine learning in this domain has largely relied on diffractograms simulated from known structures\cite{Park2017, Lee2023} or, most recently, from generated synthetic crystals\cite{Schopmans2023}. 
Models trained on datasets with simulated diffractograms have already shown strong performance in predicting phases \cite{Park2017,chenAutomatingCrystalstructurePhase2021, changProbabilisticPhaseLabeling2023}, lattice parameters\cite{Dong2021, Chitturi2021, Habershon2004, zhang2024crystallographic}, spacegroup \cite{cao2024simxrd, Schopmans2023, Oviedo2018, Park2017, Vecsei2018, Zaloga2020, Suzuki2020, Chakraborty2021,zhang2024crystallographic}, and crystallite size \cite{Dong2021, Chakraborty2021} from simulated diffractograms.
However, the performance substantially drops off when these models are applied to data originating from experiments \cite{cao2024simxrd, Schopmans2023,zhang2024crystallographic, Wang2020, Vecsei2018}. This discrepancy in performance arises due to imperfections in experimental data which are not present in diffraction patterns modeled under ideal conditions. This is discussed in more detail below.

Both labeled and unlabeled datasets of experimental powder diffractograms hold significant value for machine learning-based pXRD analysis, particularly with regard to bridging the performance gap between simulated and experimental domains. Labeled experimental data can be used to test and benchmark existing and new automated analysis approaches. This enables researchers to gauge how well a given model would perform under real-world conditions if integrated into an automated experimentation pipeline. Unlabeled experimental data enables machine learning researchers to evaluate how closely their simulations represent experimental data and modify their simulation algorithms accordingly. Unlabeled data can also find applications in transfer learning approaches to transfer model capabilities from the domain of simulated diffractograms to the domain of experimental diffractograms. While some experimental powder databases exist, their utility is limited by the fact that they are either small or not openly accessible.

\begin{figure}[!htb]
    \centering
    \includegraphics[width=1.0\linewidth]{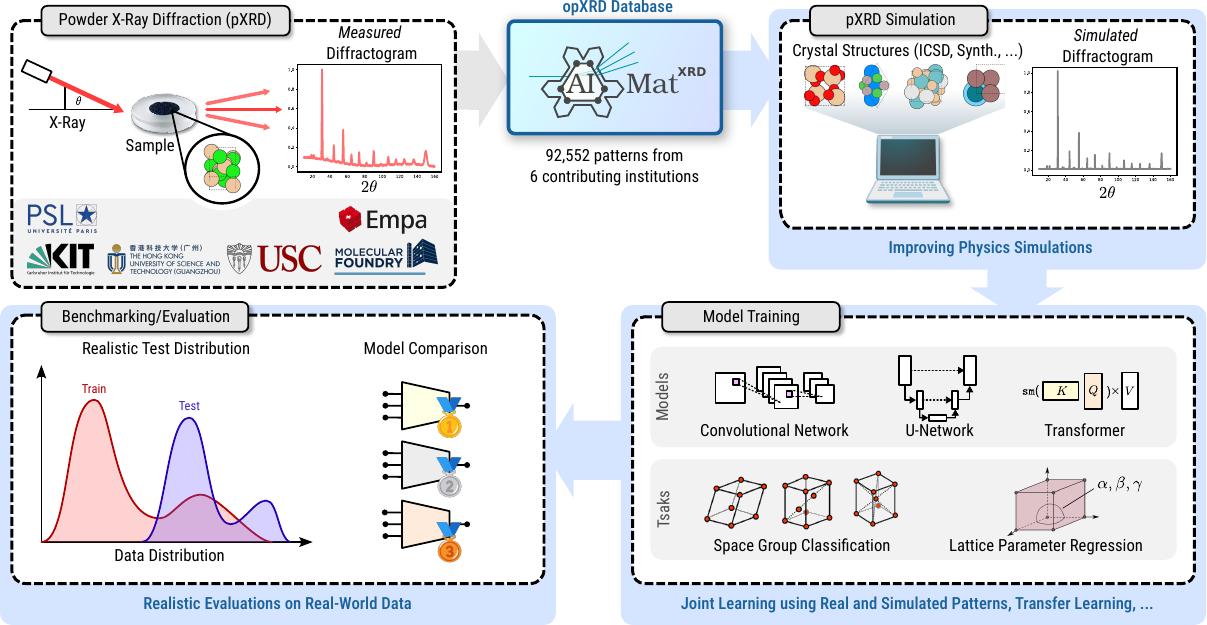}
    \caption{Experimental powder X-ray diffraction (pXRD) patterns from several contributors are collected in the \textit{opXRD} database. The proposed open-access database of experimental data aims to support each step in the pXRD-related machine learning workflow by informing better physics simulations, supplying model training data, and providing a foundation for realistic performance evaluations.}
    \label{fig:ml_uses}
\end{figure}

In this work, we introduce an open powder X-ray diffraction (opXRD) database featuring a broad range of patterns collected from experiments. With a total of $92,552$ patterns collected from 6 contributing institutions, the opXRD database exceeds the size of the previously largest database of openly accessible experimental powder diffraction data by two orders of magnitude. To the best of our knowledge, the largest database of this type is the RRUFF database, containing 1290 experimental powder diffraction patterns \cite{lafuente2015}. Larger commerical datasets such as the PDF5+\cite{GatesRector2019} and the Linus Pauling File\cite{villars2018} exist, but their utility is limited by fees and restrictive licenses. License terms of commercial datasets, such as the PDF5+ and the Linus Pauling File, prohibit or restrict the publication of models trained on their data. In contrast, the opXRD database is both free and imposes no restrictions on how its data is used. Fig.~$\eqref{fig:ml_uses}$ provides an overview of machine learning workflows enabled and supported by the opXRD database.

Of the \numpatterns patterns in the opXRD database \numlabeled patterns come with at least partial structural information of the underlying sample. Of these \numlabeled labeled patterns, more than 900 have full structural labels including atomic coordinates. This constitutes an experimental pXRD test dataset that is larger in size, richer in labels, and broader in represented experimental setups than the RRUFF database, which only provides lattice parameters as labels\cite{Armbruster2015}. However, since the majority of the opXRD database is unlabeled we also want to further discuss the uses of unlabeled data, including its role in improving pattern simulations and its application in transfer learning approaches.

The neglected effects that lead to discrepancies between simulated patterns and patterns stemming from experiments are largely known. Unaccounted effects may include preferred crystallite orientation, variations in grain size, crystal defects, the impact of temperature on the scattering process, internal stress, the non-monochromaticity of the X-ray source, and X-ray-induced fluorescence\cite{cao2024simxrd, Waseda2011, Pecharsky2023}. Additionally, varying experimental setups produce distinct powder diffraction patterns on the same sample. Features that may vary between experimental setups include the shape of diffraction peaks, the wavelength and polarization of the employed X-ray source, and the detector geometry \cite{cao2024simxrd, Waseda2011, Pecharsky2023}. The recorded scattering angles may also be slightly falsified if the sample is displaced from its intended position\cite{cao2024simxrd,hulbert2023}. As these and more neglected effects are integrated into the simulation process, real powder diffraction data can be used to evaluate how closely simulated data matches up with real data. While direct comparisons are only possible on labeled patterns, comparing the strength and prevalence of features between simulated and real data can nevertheless provide information about the fidelity of the simulation. Taking into account all neglected effects without making approximations will incur significant computational costs that will lower the size of the generated training data. A more efficient approach could be to use real experimental data to identify the effects that have the largest impact in practice and model them heuristically.

The second way in which unlabeled experimental data can serve to bridge the performance gap between simulated and experimental domains is through transfer learning. The objective of transfer learning is to transfer the capabilities of a model learned on a source domain in which labeled data is abundant to a target domain in which labeled data is sparse\cite{Zhuang2021}. In this context, the source domain is simulated powder diffraction patterns and the target domain is experimental powder diffraction patterns. Many approaches to transfer learning have been proposed, particularly in the domain of image classification \cite{Gatys2016, Ganin2015}.  These existing techniques can be adapted to facilitate transfer learning in the context of pXRD patterns. Seddiki \textit{et al}. have already successfully applied transfer learning in the domain of mass spectrometry to boost the accuracy of mass spectrum classification models\cite{Seddiki2020}. Since both mass spectrometry data and pXRD data are one-dimensional, this work demonstrates the merit of transfer learning in a setting similar to pXRD.

\begin{figure*}[!htb]
    \centering
    \includegraphics[width=\linewidth]{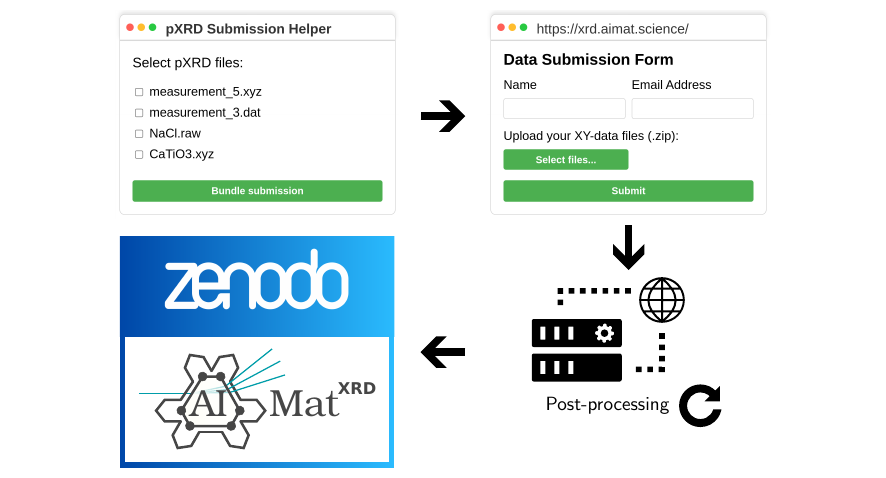}
    \caption{Overview of the data collection pipeline. Datasets are submitted using an online submission form, optionally with the help of our submission helper software. After post-processing and data homogenization, we offer the creation of a Zenodo entry for each user submission and subsequently include the submission in the opXRD database.}
    \label{fig:submission}
\end{figure*}

The opXRD database is intended as a growing, community-driven initiative. The database we present here is the first version, but we hope to further increase the database size through active engagement with the pXRD community. Our primary objective is to minimize the effort and thus the barrier to contributing experimental data to the opXRD database. Thus, we developed a program that helps to find and share data from pXRD lab computers. Users can select their most common pXRD file types, the program lists all files of that type, and users can select or deselect certain folders or files for sharing. Selected contributions will be uploaded to opXRD, processed to a common file format, and---if wanted---published on Zenodo on behalf of the contributors, before becoming part of the opXRD database. If labels are available, they can be shared with opXRD as well. Further details can be found on the opXRD website (\url{https://xrd.aimat.science/}). An overview of this process is given in Fig.~$\eqref{fig:submission}$ below.

As argued by Aranda and Kroon-Batenburg \textit{et al}.\cite{Aranda2018, Kroon-Batenburg2024}, sharing raw powder diffraction data is not only in the interest of furthering machine learning research but is also in line with open science principles. It furthers the ability of other researchers to reproduce published work and in turn, adds to the credibility of the publisher of the data. Compared to publishing data individually, publishing data on the opXRD database has the added benefit of contributing to a large, homogenous dataset with a standardized interface. This makes the data more easily accessible to other researchers and provides more value to researchers seeking large quantities of data. However, further data annotation with metadata is required to fully fulfill the FAIR data principles.

The opXRD database contains pXRD patterns from single and multiphase materials from a wide variety of material classes, including high-entropy materials, perovskites, and commercial catalysts. Some of the XRD data was collected on thin-films rather than on true powder samples, which may influence the quality of the data in regards to full structure resolution. Additionally, some of the data was collected in grazing-angle geometry rather than in the usual Bragg-Brentano geometry employed in powder diffraction.
The broad range of available experimental samples contained in the opXRD v1.0 database makes it possible to apply state-of-the-art ML approaches to the domain of pXRD analysis. We hope that the opXRD database can drive ML research in this field towards more advanced automated analysis workflows that can accelerate materials science research through ready application in high-throughput experimentation pipelines. Details of the experiments of research groups contributing to the opXRD database are discussed in Section~$\eqref{sec:our_dataset}$. A detailed description of how to acquire and use opXRD data is given in Section~$\eqref{sec:how_to_use}$, and Section~$\eqref{sec:summary_and_outlook}$ describes how further data can be contributed.

\subsubsection*{Review of machine learning-based pXRD analysis}
To showcase the need for datasets such as the one presented in this publication, we now discuss some recent approaches that apply machine learning methods to classification and regression tasks for powder diffractograms.

In 2020, Lee {\it et al.} trained a deep convolutional neural network (CNN) using simulated diffractograms based on structures from the ICSD, which is able to classify occurring phases in diffractograms of a specific compound pool \cite{Lee2020}. In 2022, they furthermore developed models based on fully convolutional neural networks and transformer encoders that predict the crystal system, the spacegroup, and other structural properties, such as the band gap \cite{Lee2022}. With their best model for the crystal system prediction on ICSD structures, they achieved a test accuracy of \SI{92.2}{\percent}. In 2017, Park {\it et al.} reached a test accuracy of roughly \SI{81}{\percent} for a CNN, which classifies space groups of simulated single-phase diffractograms \cite{Park2017}.

A regression analysis on lattice parameters within a broader framework encompassing all material classes was conducted by Chitturi {\it et al} \cite{Chitturi2021} in 2021. They developed a distinct CNN for each crystal system, utilizing a merged dataset from both the ICSD and the Cambridge Structural Database, and managed to achieve a mean absolute percentage error of about \SI{10}{\percent} for the lattice lengths, although they encountered difficulties in accurately predicting angles.
In 2024, Zhang {\it et al.} introduced a convolutional self-attention neural network trained on simulated patterns to classify crystal types \cite{zhang2024crystallographic}. Their model was tested on 23,073 unary, binary, and ternary inorganic crystal structures sourced from the COD. The study observed a noticeable performance drop when the pre-trained model was applied to real experimental patterns as opposed to simulated data. However, their recent work \cite{cao2024simxrd} proposes using convolutional peak descriptors that consider the detector's geometry, which reduces the performance gap in their benchmark tests.

Neural networks trained purely on experimental diffractograms can perform well when the range of samples is narrow and the data is collected only on a single machine \cite{Lee2023, hattrick-simpers2021}. However, in a more general setting with a wide range of investigated samples and employed diffractometers training neural networks purely on experimental diffractograms becomes infeasible. This is because of the limited availability of labeled experimental diffractograms relative to the scope of the task. However, in 2023, Salgado {\it et al.} \cite{Salgado2023} showed that adding a fraction of experimental patterns to a simulated training dataset improves the performance on unseen experimental patterns. They used \SI{50}{\percent} of the experimental patterns contained in the RRUFF database and added those to their large simulated training set. Then they tested their model's performance on the other half of the RRUFF database and achieved a performance increase in the 230-way spacegroup classification accuracy of \num{11} percentage points compared to the same model only trained on simulated patterns.

In 2024, Schuetzke {\it et al.} trained a classifier to classify if a diffractogram stems from an amorphous, single-phase, or multi-phase sample \cite{Schuetzke2024}. Due to the lack of experimental pXRDs, they built a pipeline to augment simulated diffractograms of a reference structure by, among other things, slightly varying the underlying crystal lattice. For spinel structures, they reported an accuracy of \SI{100}{\percent} but they also proved that their approach can be transferred to other datasets.

In 2023, Schopmans {\it et al.} presented an approach to generate synthetic crystal structures and their corresponding pXRD patterns on the fly during the training process \cite{Schopmans2023}. This approach defeats the issue of a limited dataset size, which limits the depth of neural networks that can be trained. However, the accuracy dropped substantially when we applied our space group classification model to experimental patterns from the RRUFF database. Augmenting our simulated patterns with background, noise, and impurities helps to bring simulated diffractograms closer to experimental ones, making models trained on them more performant on experimental diffractograms. However, this augmentation process could be improved by incorporating background and noise statistics from a broader experimental pXRD database, such as the one presented in this publication.

It becomes apparent that the more general the task is, the more challenging the transfer to experimental data becomes. For example, the space group classification task across all material systems is very general. Therefore, transferring it to the application on experimental diffraction patterns is difficult. \cite{Schopmans2023, Lee2022, Vecsei2018} On the other hand, there are some successful approaches that also work well on experimental data, but those are mostly methods that do phase determination in a limited compound space, making the task less complex \cite{Schuetzke2024, Lee2020}. 

The current volume of experimental pXRD patterns is insufficient to effectively train ML models, highlighting an urgent need for a comprehensive experimental pXRD database. The most advanced ML models currently are trained on approximately $10^5 - 10^6$ simulated diffractograms \cite{Salgado2023, Schopmans2023}. This is, to the best of our knowledge, two orders of magnitude larger than the largest currently curated experimental dataset, the PDF-5+ with approximately $2\cdot 10^4$ experimental patterns. It is even one order of magnitude larger than the approximately $10^5$ unlabeled diffractograms in the initial version of the opXRD dataset we present here.

To make ML-based pXRD data identification practical for experimental use and automate structure prediction despite lacking experimental training data two key approaches are essential. First, developing more sophisticated simulation methods to better approximate experimental patterns\cite{cao2024simxrd} by using statistics from experimental diffractograms. Second, creating an experimental database that enables transfer learning to bridge the gap between simulated and real-world data. For both of these steps, the development of opXRD is particularly significant, as it will provide a comprehensive experimental benchmark for the community, allowing fair comparison of baseline models and accurate evaluation of their applicability in real experimental situations.

\section{Existing datasets}
To contextualize opXRD within the current environment of experimental powder diffraction data, the list below provides an overview of the largest crystal structure databases that offer access to experimental powder diffraction data. For an overview of these databases refer to Tab.~$\eqref{tab:exp_databases}$ below. 

\begin{table}[!htb]
\centering
\caption{\footnotesize Overview of experimental powder diffraction databases: The column ``O.A.'' indicates whether or not the database is open-access. The availability of the chemical composition, spacegroups, lattice parameters, and atomic coordinates of the underlying samples are indicated by the columns ``Comp.'', ``Spg.'', ``Lattice'' and ``Atom coords.'', respectively.}
\label{tab:exp_databases}
\scalebox{0.82}{
\begin{adjustbox}{center}
\begin{tabular}{@{}llllllll@{}}
\toprule
\textbf{Name} & \textbf{No. patterns} & \textbf{O.A.} & \textbf{Comp.} & \textbf{Spg.} & \textbf{Lattice} & \textbf{Atom coords.} & \textbf{Year est.} \\
\midrule
Linus Pauling file                     & 21,700                    & \xmark          & \cmark        & \cmark        & \cmark        & \cmark          & 2002            \\
Powder Diffraction File\tablefootnote{The PDF lists the Material Platform for Data Science (MPDS) as a database source. Since the MPDS is hosted by the Pauling File project, there is likely significant overlap in the experimental patterns available in the PDF and the Linus Pauling File.} & 20,800 & \xmark & \cmark & \cmark & \cmark & \partialcheck{52} & 1941 \\
RRUFF                                  & 1290                      & \cmark          & \cmark        & \cmark        & \cmark        & \xmark          & 2006            \\
Crystallography Open Database          & 1052                      & \cmark          & \cmark        & \cmark        & \partialcheck{85}        & \partialcheck{85}          & 2003            \\
PowBase                                & 169                       & \cmark          & \cmark        & \xmark        & \xmark        & \xmark          & 1999            \\
\bottomrule
\end{tabular}
\end{adjustbox}}
\end{table}

\textbf{Linus Pauling File}:\cite{PaulingWeb} The Linus Pauling File is a largely commercial crystal structure database published and maintained by the Pauling File project \cite{villars2018}. It is currently distributed as Pearson Crystal data \cite{PearsonWeb} and the Materials Platform for Data Science (MPDS)\cite{MPDSWeb}. The database, first published in 2002, currently contains more than 534,000 crystal structures\cite{MPDSWeb} and 21,700 corresponding experimental powder diffraction patterns\cite{PearsonWeb}. 
This makes the Pauling file, to the best of our knowledge, the largest collection of experimental powder diffraction data available to researchers. As of November 2024, Pearson's crystal data is available to researchers through a purchase of a one-year license starting at a price point of \qty{2200}{\myeuro}\cite{PearsonBuy}. The MPDS is partially open, with the open portion of the MPDS data accessible through a web interface\cite{MPDSWeb}. API access to the full MPDS can be purchased through a one-year license starting at \qty{9500}{\myeuro}\cite{MPDSBuy}. We asked the Pauling File project whether the experimental powder diffraction data is accessible through the MPDS API. The Pauling File project responded that this data is not currently provided through the API, but could be offered in the future at the request of customers.

\textbf{Powder Diffraction File:} \cite{PDFWeb} The Powder Diffraction File (PDF), published and maintained by the International Center for Diffraction Data (ICDD), is a large collection of materials with accompanying powder diffraction data first published in 1941\cite{GatesRector2019}. According to the ICDD the latest release of the PDF, the PDF5+, contains over a million materials with accompanying powder diffraction data. However, since most of these powder diffraction patterns are simulated we asked the ICDD about the number of experimental diffraction patterns in the PDF5+. We were told that 20,800 of the patterns in the PDF5+ stemmed from experiments and that 10,954 of these patterns were accompanied by the atomic coordinates of the underlying structures. Since the PDF5+ lists the MPDS as a database source, there is likely a significant overlap in the experimental patterns found in the PDF5+ and those found in the Pauling file. Currently, the PDF5+ is available to researchers through a purchase of a one-year license starting at a price point of \$6265. However, the ICDD does not allow researchers to train machine learning models on PDF5+ data, regardless of whether the resulting models are published \cite{PDFLicenseWeb}.

\textbf{RRUFF}: \cite{RRUFFWeb} The RRUFF Mineral Database, first published in 2006, provides detailed information on minerals, including their chemical compositions, crystallography, and spectroscopic data \cite{lafuente2015}. Managed by the University of Arizona, it was created to serve as a public repository for mineral identification and research. It contains \num{1290} powder diffraction patterns stemming from experiments each labeled with the lattice parameters and composition of the underlying structures. The RRUFF data is openly accessible on its official website \cite{RRUFFWeb}.

\textbf{Crystallography Open Database:} \cite{CODWeb} The Crystallography Open Database (COD) is an open-access collection of crystal structures founded in 2003\cite{Graulis2009cod}. It currently provides over 500,000 crystal structures. Of these files, 1052 contains the experimental powder diffraction data that was used to determine the underlying crystal structures of the investigated samples. Hence, the experimental powder diffraction data contained in the COD is mostly labeled with the full crystal structure information. The data is openly accessible in the form of .cif files on the official COD website\cite{CODWeb}.

\textbf{PowBase:} \cite{PowBaseWeb} PowBase is a database of 169 mostly unlabeled experimental powder diffraction patterns collected and maintained by crystallography researcher Armel Le Bail starting in 1999. PowBase is an initiative suggested in the Structure Determination by Powder Diffractometry (SDPD) mailing list which was co-maintained by Le Bail. The COD is another community initiative that grew out of this mailing list. As of March 2025, all 169 patterns are still freely available for download on the official website \cite{PowBaseWeb}.

There is also publicly available powder diffraction data uploaded to datasets on Zenodo. However, this data is split into disparate entries that typically only contain the work of a single research project. Additionally, extracting powder diffraction data at scale is hindered by the fact that the data is often given in plain text files in non-standardized formats, which are difficult to parse automatically. We are currently planning a systematic large-scale extraction of powder diffraction data from databases like Zenodo with the help of a large language model. This data will be included in a future release of the opXRD database.

While not strictly speaking a powder diffraction database, the High-Throughput Experimental Materials Database (HTEM) by the National Renewable Energy Laboratory (NREL) is a valuable source of X-ray diffraction data \cite{zakutayev2018}. Currently, the HTEM database contains 65,779 thin-film samples with corresponding X-ray diffraction data\cite{HTEMWeb}. Each database entry includes the elemental composition of the underlying sample but does not provide any information on its structure. HTEM data is open-access and can be downloaded through an API provided by NREL.

Aside from the databases mentioned above, we have also investigated several other crystal structure resources in search of experimental powder diffraction data. Crystal structure resources that were investigated but not found to contain any appreciable amount of publicly available experimental powder diffraction data include the Inorganic Crystal Structure Database \cite{ICSDWeb}, the Cambridge Structural Database \cite{CambridgeWeb}, the Materials Project database \cite{MatProjWeb}, the Crystallographic and Crystallochemical Database \cite{CrystallochemicalWeb}, the Bilbao Incommensurate Crystal Structure Database \cite{BilbaoWeb}, the Mineralogy Database \cite{MineralogyWeb}, the IUCr Raw data letters \cite{IUCrWeb}, the U.S. Naval Research Laboratory Crystal Lattice-Structures \cite{NRLWeb}, the Athena Mineral database \cite{AthenaWeb} and the Protein data bank\cite{PDBWeb}. The lack of experimental powder diffraction data in these databases is to be expected as most structure solutions are achieved through single-crystal diffraction.

%\pagebreak
\section{opXRD database}\label{sec:our_dataset}
In collaboration with several other research institutions, we have collected a database of $\numpatterns$ experimental patterns of which 2179 are at least partially labeled with structural information of the underlying material. The following research institutions contributed data to the opXRD database: The French National Centre for Scientific Research (CNRS), Hong Kong University of Science and Technology (Guangzhou) (HKUST), University of Southern California (USC), Lawrence Berkeley National Laboratory (LBNL), Empa–Swiss Federal Laboratories for Materials Science and Technology (EMPA) and the Karlsruhe Institute of Technology (KIT). Tab.~$\eqref{tab:merged}$ provides an overview of the contributions of each institution. We filtered the submitted datasets to exclude patterns with invalid features such as only one unique recorded angle, negative angles, less than 50 recorded angles total, or all intensities being zero.

\begin{table}[!htb]
\centering
\caption{\footnotesize Overview of the contributions to the opXRD database: The availability of the chemical composition, spacegroups, lattice parameters, and atomic coordinates of the underlying samples are indicated by the columns ``Comp.'', ``Spg.'', ``Lattice'' and ``Atom coords.'' respectively.}
\label{tab:merged}
\scalebox{0.735}{
\begin{adjustbox}{center}
\begin{tabular}{@{}lllllll@{}}
%\begin{tabular}{@{}llccccl@{}}
\toprule
\textbf{Institution} & \textbf{No. patterns} & \ \textbf{Comp.} & \ \ \textbf{Spg.} & \textbf{Lattice} & \textbf{Atom coords.} & \textbf{Research Project} \\
\midrule
CNRS        &  1052          & \ \cmark   & \partialcheck{85}   & \cmark   &  \ \ \quad \partialcheck{85}   & Diffraction data extracted from the COD \\ 
USC         &  338          & \ \cmark   & \cmark   & \partialcheck{90}   & \ \ \quad \xmark   & Study of CuNi and CuAl alloys          \\
HKUST(GZ)   &  520          & \ \partialcheck{4}   & \partialcheck{4}   & \partialcheck{4}   & \ \ \quad \partialcheck{4}   & Phase identification dataset  \\ 
EMPA        &  770          &\ \cmark   & \partialcheck{63}   & \xmark   & \ \ \quad \xmark   & Metal halide perovskites, Zn-V-N libraries \\ 
INT         &  19,796        &\ \xmark   & \xmark   & \xmark   & \ \ \quad \xmark   & Compilation of various projects        \\ 
IKFT        &  64           &\ \xmark   & \xmark   & \xmark   & \ \ \quad \xmark   & Commercial catalysts, metals, metal oxides   \\ 
LBNL        &  70,012          &\ \xmark   & \xmark   & \xmark   & \ \ \quad \xmark   & Perovskites precursors, Mn-Sb-O system  \\ 
\bottomrule
\end{tabular}
\end{adjustbox}}
\end{table}

\begin{figure}[!htb]
    \centering\includegraphics[width=0.85\linewidth]{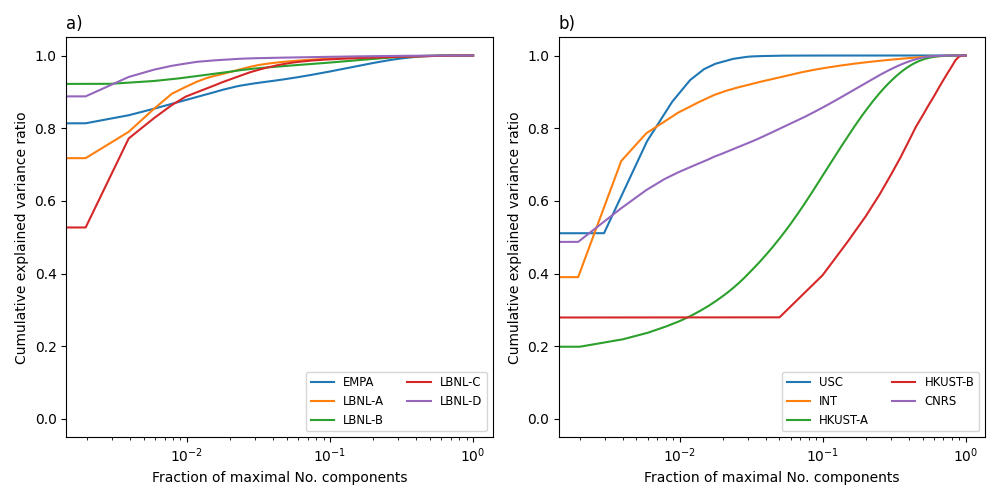}
    \caption{Explained variance ratio over the fraction of the maximum number of components for each dataset contributed to the opXRD database. Here the maximal No. components refers to $N_{\text{max}}$ as defined in equation $\eqref{eq:nmax}$. Datasets contributed by the same institution are labeled alphabetically in the order in which they are described in the texts towards the end of this section.}
    \label{fig:components}
\end{figure}

The variance of the data was analyzed using principal component analysis (PCA). PCA can be applied to datasets $X \subset \mathbb{R}^N$ to reduce the number of components needed to describe points $p \in X$ up to some tolerance in lost accuracy. In the context of PCA, the cumulative explained variance ratio is a measure of how much of the variance in the dataset $X$ can be explained using a given number of components. For a rigorous definition of PCA and the explained variance ratio, we refer to the literature \cite{Jolliffe2016}. Here, PCA was performed on datasets of X-ray diffraction patterns. These datasets $X$ are subsets of $\mathbb{R}^{N}$ with $N=512$ since each pattern $p \in X$ was standardized to have 512 intensity values spread out evenly from $0\degree$ to $180\degree$ using zero padding and interpolation with cubic splines. Hence the maximal components that could be needed to describe a dataset of diffraction data in this context is $N=512$. However, the maximal number of components is even lower for datasets that contain less than 512 patterns. In this case, the maximal number of components is equal to the number of patterns in the dataset since each pattern can add at most one degree of freedom to the dataset $X \in \mathbb{R}^N$. Hence the maximum number of components $N_{\text{max}}$ of a pattern dataset $X$ is given as follows:
\begin{align}
    N_{\text{max}} = \min(N_{\text{values}}, N_{\text{patterns}}).
      \label{eq:nmax}
\end{align}
Here $N_{\text{values}} =512$ is the number of recorded intensity values per pattern and $N_{\text{patterns}}$ is the number of patterns in the dataset $X$. Fig~$\eqref{fig:components}$ below shows the cumulative explained variance ratio over the fraction of maximal No. components $N_{\text{max}}$ as defined above. In this figure, a faster convergence of the cumulative variance ratio towards one indicates that the patterns in this data are relatively similar. The degree of variation between the patterns is different for each contribution. For example, the CRNS and the HKUST contributions each are collections that encompass many research projects over a large period of time and thus exhibit a high degree of variability between individual patterns. In contrast, the contributions by USC and LBNL contain many very similar patterns. The patterns in the USC dataset are similar because the underlying samples are all variations of \ce{CuNi} and \ce{CuAl} alloys. The patterns submitted by LBNL are similar because they stem from in-situ recordings where several hundred or several thousand patterns were collected over time per sample while they were undergoing physical conversion processes.

\begin{figure}[!htb]
    \centering
    \includegraphics[width=0.8\linewidth]{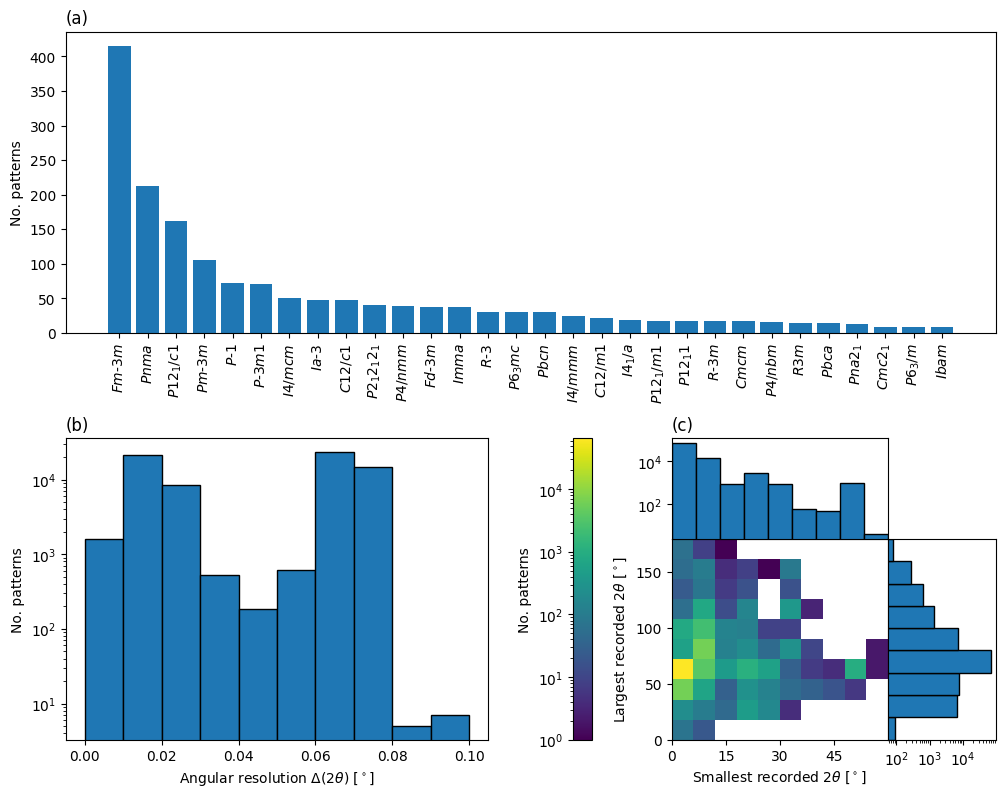}
    \caption{Histograms detailing the distribution of pattern and structure properties in the opXRD database: a) distribution of spacegroups present in labeled data; b) distribution of angular resolution in all data; c) distribution of smallest and largest recorded $2\theta$ values for all data.}
    \label{fig:histograms}
\end{figure}

Fig.~$\eqref{fig:histograms}$ provides an overview of the distributions of pattern and structure properties in the opXRD database. Nearly all patterns have an angular resolution smaller than $\Delta(2\theta) = 0.1 ^\circ$. Here the angular resolution is defined as the range of recorded angles divided by the number of recorded intensity values along that range. For most patterns, the lowest recorded angle is smaller than $30 ^\circ$ and the highest recorded angle is smaller than $120 ^\circ$. The start-to-end angle distribution reveals that all diffractograms start in a narrow window between $0\degree$ and approximately $50\degree$, while they end between $50\degree$ and $150\degree$, with the majority of patterns going from $0\degree$ to approximately $70\degree$. Unlike most ML approaches using synthetic data over the full angle range with fixed resolution, the opXRD dataset has a strongly varying angle range and resolution. Hence, working with this data requires additional pre-processing methods such as padding and interpolation, or more flexible ML models beyond standard CNNs.

In the following, we will describe the datasets contributed by each of the collaborating research groups and institutions. Each paragraph includes a description of the investigated materials and how X-ray diffraction data was collected. If applicable, the presence of thin-film samples or atypical diffraction geometries is indicated. Most data was collected using \ce{Cu} radiaton sources which has a $K_{\alpha1}$ wavelength of $\lambda=1.54056\text{\AA}$ and a $K_{\alpha2}$ wavelength of $\lambda=1.54439\text{\AA}$.

%% Paragraph by FX Coudert and Arthur Hardiagon
\subsubsection*{Institut de Recherche de Chimie Paris, CNRS}

Experimental pXRD data was extracted from the Crystallography Open Database (COD)\cite{Grazulis2009, Vaitkus2023}. The COD is, to our knowledge, the largest open-access collection of experimental crystal structures of organic, inorganic, and metal-organic compounds and minerals, containing more than 500,000 entries. The data in the COD are placed in the public domain and licensed under the CC0 License. Of the entire COD database 5432 structures contained at least one tag from the {CIF\_POW} dictionary, i.e., a tag relating to powder diffraction studies. These 5432 structures only account for 1\% of the total COD database, but this is to be expected since most crystal structures are resolved from single-crystal diffraction. Of these 5432 files, most contained only metadata related to the powder diffraction experiment, but did not include the raw data of the pattern itself. We could extract raw experimental pXRD patterns from 1052 files in total, after curation of a small number of files with clearly invalid data.

The pXRD data from the COD database are of high quality, with a median resolution of $\Delta(2\theta) = 0.013\degree$ and an average number of 9190 points measured per pattern. They span a wide chemical space, including organic, inorganic, and hybrid structures, and 75 different elements of the periodic table.

% Paragraph by Bin Cao and Tong-yi Zhang
\subsubsection*{Guangzhou Municipal Key Laboratory of Materials Informatics, HKUST(GZ)}

Two datasets were contributed to the opXRD database. The first dataset (HKUST-A) is a selected subset of a small-scale experimental powder X-ray database developed over the past two years, called the X-Ray Phase Identification Public Experimental Dataset (XRed) (\url{https://github.com/WPEM/XRED}). The primary goal of XRed is to support the advancement of intelligent phase identification technology by providing a foundation for data collection in future large-scale machine learning applications. XRed primarily focuses on metal and metal-oxide particles, with data collected using diffractometers such as the Empyrean 3.0, Aeris, and Bruker D8 Advance, all employing \ce{Cu} X-ray sources. The dataset HKUST-A contains 21 pXRD patterns each labeled with a corresponding CIF file that documents the refined structure. Data are categorized by elemental systems and include original experimental files, spanning single-phase to five-phase mixtures, as well as mixtures designed for various research tasks.

In addition to XRed, the opXRD database integrates an experimental dataset composed of powder diffraction data sourced from open-access publications and collaborating institutions (HKUST-B). These institutions have provided the data with full authorization for research purposes. Compared to XRed, this dataset offers broader chemical element coverage, encompassing ionic, atomic, and metallic crystals. It is also larger, containing 499 entries. However, unlike XRed, these data entries are not accompanied by CIF files.

% Paragraph by Alexander Wieczorek and Sebastiann Siol
\subsubsection*{Laboratory for Surface Science and Coating Technologies, Empa}

Combinatorial Zn–V–N libraries were synthesized using radio-frequency co-sputtering of Zn and V in a mixed Ar and \ce{N2} plasma. An orthogonal deposition temperature and composition gradient was created, resulting in a deposition temperature of $220\degree$C for samples 1~–~9 and $114\degree$C for samples 37~–~45. The composition for each sample was determined using X-ray fluorescence (XRF) spectroscopy which was further calibrated through Rutherford backscattering spectroscopy (RBS) based on selected samples. The newly identified and isolated semiconductor \ce{Zn2VN3} was identified to exhibit a cation-disordered wurtzite structure as verified by additional GI-XRD and SAED measurements\cite{Zhuk2021}.

Tin halide perovskites were deposited using single-step spin-coating as reported elsewhere\cite{Wieczorek2023}. Methylammonium lead iodide libraries with varying degrees of residual \ce{PbI2} were deposited using a two-step procedure involving both thermal evaporation of \ce{PbI2} and subsequent spin-coating of a methylammonium solution. The relative phase fractions were quantified using supplementary azimuthal angle scans coupled with structural factors and geometrical factors as reported elsewhere\cite{Wieczorek2024}. Fully inorganic lead perovskite libraries were prepared using thermal co-evaporation of lead and cesium halide salts. All metal halide perovskite libraries were measured within a custom-made X-Ray transparent inert-gas dome, resulting in the presence of minor additional features within the $\theta=19 \text{–}31\degree$ range. For all combinatorial libraries where any phases are specified, the complete set of phases is reported in the metadata.

XRD data was measured using a Bruker D8 Discover equipped with a Cu radiation source in a Bragg-Brentano geometry. For the reported data sets the instrument was equipped with a Goebel mirror effectively removing the Cu~K$\beta$ radiation. The data set originates from the combinatorial exploration of the Zn–V–N compositional space, as well as data gathered from multiple research activities on more established metal halide perovskite semiconductors. All data was collected from thin films deposited on borosilicate glass. The Zn–V–N films showed some preferential out-of-plane orientation, while for the perovskites the preferential orientation was minimal, resulting in the presence of all reflections.

%% Paragraph by Ben Breitung
\subsubsection*{Institute of Nanotechnology, KIT}

X-ray diffraction data was collected from a wide range of research projects conducted at the Institute of Nanotechnology over the past 10 years. A major part of the research focused on high-entropy materials, which involved incorporating many different elements into single-phase structures, leading to peak shifts or phase separations. Most of those multi-component complex materials appeared in various structures, including rock-salt, spinel, fluorite, perovskite, and delafossite. The samples were prepared either in powder or in bulk form; therefore, powder XRD was performed on samples with adjusted height. The samples were prepared using various synthesis techniques, mostly solid-state or wet chemical syntheses, to obtain the desired structures. Consequently, particle size and crystallinity varied significantly. The sample set also includes samples that were not successfully measured or where phases could not be identified.

The X-ray diffraction data were collected on a Bruker D8 Advance using a Cu radiation source or a STOE Stadi P diffractometer equipped with a Ga-jet X-ray source. The samples were initially recorded for various research projects over the last ten years and were measured with different step sizes, times per step, and over different angle ranges, but all using \ce{Cu} $K_\alpha$ or \ce{Ga} $K_\beta$ radiation. The samples mostly contained transition metal oxides, sulfides, and fluorides. To improve statistics, the samples were rotated during the entire measurement. Some air-sensitive samples were measured using a transparent polymer dome for protection. This dome led to increased background noise over the first $20\degree$ and slightly decreased pattern resolution.

% Paragraph by Moritz Wolf 
\subsubsection*{Institute of Catalysis Research and Technology, KIT}

A variety of samples were analyzed including commercial catalysts, bulk reference materials, porous metal oxide particles, and nanoparticles. The latter were synthesized via the surfactant-free benzyl alcohol route \cite{Wolf2019, Wolf2018}. The cobalt oxide (\ce{CoO} or \ce{Co3O4}) and cerium oxide (\ce{CeO2}) nanoparticles were in the size range of $4-16 \ \si{nm}$ according to the Scherrer equation. A series of porous \ce{Al2O3} materials, which were prepared by calcination of boehmite (\ce{AlOOH}) at various temperatures, represents crystalline samples with limited long-range structure and various contributions of \ce{Al2O3} polymorphs.

X-ray diffraction (XRD) was conducted with an X’Pert Pro MPD (Panalytical) in Bragg-Brentano geometry using a \ce{Cu} X-ray source. The patterns were acquired in the $2\theta$ range of $5-80\degree$ with a step size of $0.016711\degree$ or $0.033420\degree$ and a total acquisition time of 40 to 120 min. This study has been carried out with the support of Angelina Barthelmeß, Elisabeth Herzinger, and Henning Hinrichs.

% Paragraph by Tim Kodalle
\subsubsection*{Molecular Foundry Division \& Advanced Light Source \& Chemical Sciences Division, LBNL}

In total four different datasets were collected. The first dataset (LBNL-A) was collected from spin-coating and annealing triple-cation metal-halide perovskite precursor solutions with the composition \ce{Cs}$_{0.05}$(MA$_{0.23}$FA$_{0.77}$)\ce{Pb}$_{1.1}$(\ce{I}$_{0.77}$\ce{Br}$_{0.23}$)$_{3}$ onto various substrates. Here, MA stands for Methylammonium and FA stands for Formamidinium. The substrates onto which these solutions were coated include glass, which is amorphous, and $\ce{GaAs}$ wafers, which are single crystalline. Other substrates were stacks of glass/indium tin oxide, stacks of \ce{GaAs}/CIGS, and stacks of glass/CIGS. Here, CIGS stands for a stack of \ce{Mo}, \ce{Cu}(\ce{In}, \ce{Ga})\ce{Se2},\ce{Cds} and \ce{ZnO}. Some of the substrates were additionally covered with a self-assembling monolayer of MeO-2PACz. The $\ce{GaAs}$ substrates were prepared by Dr. Jiro Nishinaga from the National Institute of Advanced Industrial Science and Technology (AIST) in Japan \cite{nishinaga2018} and the glass/CIGS substrates by Dr. Christian Kaufmann and his team at Helmholtz-Zentrum Berlin (HZB) in Germany \cite{heinemann2017}. Data collection was performed in situ during thin-film deposition using a custom-made spin-coating and annealing stage \cite{song2019}. 

A second dataset (LBNL-B) was collected from spin-coating metal-halide perovskite precursor solutions with varying compositions of \ce{MAPb(I_{1-x}Br_x)3} spin-coated onto glass substrates. Here, MA = Methylammonium and x = 0, 0.33, 0.5, 0.67, 1. The substrates were preheated to different temperatures including $30\degree $C, $50\degree $C, $70\degree $C, and $90\degree $C, and the spin-coating process was performed at a constant temperature on the preheated substrates. For both datasets, diffraction data were continuously measured during spin-coating, chemical induction of crystallization, and annealing of the samples, at $100\degree$C and $110\degree$C respectively. The diffraction data was recorded with a frequency of about $0.56 
 \ 1/\si{s}$ and $0.54 \ 1/\si{s}$. Each in situ measurement consisted of about 500 to 1000 individual diffractograms. Depending on the substrate, each series of diffractograms shows an evolution from substrate only to a combination of polycrystalline perovskite, $\ce{PbI2}$ and substrate via several intermediate phases.

For these two datasets, experimental XRD data were collected at beamline 12.3.2 of the Advanced Light Source, the synchrotron at Lawrence Berkeley National Laboratory. The data were collected using a photon energy of 10 $\si{keV}$ ($\lambda = 1.23984 \text{\AA}$), selected using a Si(111) monochromator. Measurements were taken in grazing incidence geometry, i.e. using a beam incidence angle of $1\degree$. Two-dimensional diffraction images were recorded using a Dectris Pilatus 1M area detector at an angle between $34\degree$ and $36\degree$ with a sample-to-detector distance of roughly $190 \ \si{mm}$. The two-dimensional data were calibrated using an Al$_{2}$O$_{3}$ calibration standard and integrated along the azimuthal angle.

% Paragraph by Gregoire Heymans

A third dataset (LBNL-C) was collected by observing the phase evolution of an Mn-Sb-O system with varying annealing temperatures. The temperatures used to analyze the crystal structure of the Mn-Sb-O system were chosen depending on the number of phase transitions appearing for a certain temperature range. Few changes in the crystal structure appear between room temperature and 300$\degree$C and phase transitions appeared from 300$\degree$C until 850$\degree$C. No phase transition appeared when cooling down. Therefore, the crystal structure was measured every 100$\degree$C between room temperature and 300$\degree$C; every 50$\degree$C between 300$\degree$C and 850$\degree$C; and every 200$\degree$C when cooling down. The heating and cooling rates were fixed for all the experiments at 50$\degree$C/min and the holding time was fixed to 2 min.

This data was collected using the \textit{in situ} Rigaku-SmartLab3kW diffractometer. This tool operates with SmartLab Studio II software, which can measure the X-ray diffraction during the annealing process. This enables directly showing all the phase transitions when annealing in various atmospheres such as \ce{O2}, \ce{Ar}, and \ce{NH3}. Phase transitions are analyzed with the \textit{in situ} XRD tool up to 850$\degree$C in this work. Most of the \textit{in situ} experiments were performed under an air-like 20\% \ce{O2} and 80\% \ce{Ar} environment is chosen (\ce{Ar} flow: 50 \si{sccm}, \ce{O2} flow: 10 \si{sccm}). When a 100\% Ar environment is fixed, an Ar flow of 60 \si{sccm} is input. The Bragg-Brentano (BB) mode is preferred in terms of geometry because it is more adapted in the analysis of scarce phases such as  \ce{MnSb2O6} rutile. The angular step used in the recording was 0.01$\degree$ and the scanning rate was 10$\degree$/min. 

% Paragraph by Mriganka Singh
A fourth dataset (LBNL-D) was collected from a two-step spin-coating process using metal-organic frameworks (MOFs) in perovskite precursor solutions, deposited onto glass substrates. In the first step, a nanoscale thiol-functionalized UiO-66-type Zr-based MOF (\ce{UiO}-66-\ce{(SH)2)} was added to the \ce{PbI2} precursor. This was followed by the deposition of an organic mixture solution containing \ce{FAI}, \ce{MACl}, and \ce{MABr} in the second step. The incorporation of MOFs aids in suppressing perovskite vacancy defects, thereby enhancing device stability and efficiency. To further investigate the influence of \ce{UiO}-66-\ce{(SH)2)} on perovskite thin-film formation during the annealing process, a time-resolved GIWAXS experiment was conducted. The measurements were performed using a setup similar to that of LBNL-A and B.

\section{Usage}\label{sec:how_to_use}
The opXRD database is hosted on Zenodo (\url{https://zenodo.org/records/14254270}) and can be downloaded by any user without any barriers or restrictions.

Next to the availability of the opXRD dataset on Zenodo, we also provide a Python library ``opxrd'' to easily download and interface with the dataset. The instructions for how to install this library can be found in the repository associated with the library. The repository to this library is located at \url{https://github.com/aimat-lab/opxrd}. The opxrd library includes options for data-loading, standardization, plotting, and the conversion to \emph{PyTorch} tensors. We provide a Jupyter Notebook (\url{https://colab.research.google.com/github/aimat-lab/opXRD/blob/main/opxrd/usage.ipynb})
that showcases these functionalities in more detail. This notebook also illustrates how to interface with the opXRD database through Python. 

\section{Summary and Outlook}\label{sec:summary_and_outlook}
With the opXRD database, a curation of $\numpatterns$ unlabeled and 2179 at least partially labeled experimental powder X-ray diffraction patterns from a wide range of different materials systems, we provide the largest currently available source of experimental XRD patterns. With this, we address the need for experimental data that arises when developing algorithms and analysis tools for pXRD data, both based on machine learning and classical approaches. The data can be used for the actual method development and for testing. Our dataset is a valuable and so far missing resource to drive further developments in the automated analysis of XRD data.

% How to contribute
Rather than a finished project, the opXRD database is an ongoing effort to collect experimental powder XRD data. We invite everyone who is working in the area of experimental powder XRD to submit it to the dataset, in order to further improve the utility of the dataset and thus aid further developments in this field. Our submission page (\url{https://xrd.aimat.science/}) and submission helper software will be kept available to collect more data. We will keep updating and maintaining the dataset with new incoming submissions.

\subsection*{Data availability}
The opXRD database is available on Zenodo at \url{https://zenodo.org/records/14254270}. It is published under the Creative Commons Attribution 4.0 International license. It can be downloaded by any user without any barriers or restrictions. For further details, please refer to Section~$\eqref{sec:how_to_use}$.

\subsection*{Conflicts of interest}
There are no conflicts of interest to declare.

\subsection*{Acknowledgements}
H.S. acknowledges financial support by the German Research Foundation (DFG) through the Research Training Group 2450 “Tailored Scale-Bridging Approaches to Computational Nanoscience”. P.F. and D.H. acknowledge support by the Federal Ministry of Education and Research (BMBF) under Grant No. 01DM21001B (German-Canadian Materials Acceleration Center). J.Oe. and P.F. acknowledge financial support from the Helmholtz Foundation Model Initiative within Project "SOL-AI". Part of this work was funded under the France 2030 framework by Agence Nationale de la Recherche (project ANR-22-PEXD-0009 of PEPR DIADEM). Work at the Molecular Foundry was supported by the Office of Science, Office of Basic Energy Sciences, of the U.S. Department of Energy under Contract No. DE-AC02-05CH11231. Work at the Advanced Light Source (ALS) was done at beamline 12.3.2. The ALS is a DOE Office of Science User Facility under contract no. DE-AC02-05CH11231. The development of the online phase identification platform is supported by the Guangzhou-HKUST(GZ) Joint Funding Program (No. 2023A03J0003). Work by the USC group was supported by the National Science Foundation (NSF) grant numbers DMR-2227178 and OISE-2106597. M.W. acknowledges funding by the Helmholtz Research Program “Materials and Technologies for the Energy Transition (MTET), Topic 3: Chemical Energy Carriers". Work by the Empa group was supported by the Strategic Focus Area–Advanced Manufacturing (SFA–AM) through the project Advancing manufacturability of hybrid organic–inorganic semiconductors for large area optoelectronics (AMYS) as well as the Empa internal research call 2020. We thank BWCloud, funded by the Ministry of Science, Research and Arts Baden-Württemberg, for providing cloud server infrastructure.

\printnomenclature

\bibliographystyle{bibstyle/bibstyle.bst}
\bibliography{ms}

\setcounter{section}{0}
\renewcommand{\thesection}{S\arabic{section}}
\setcounter{figure}{0}
\renewcommand{\thefigure}{S\arabic{figure}}
\setcounter{table}{0}
\renewcommand{\thetable}{S\arabic{table}}

\end{document}

\maketitle
\renewcommand{\thesection}{S\arabic{section}}

\section{Description of opXRD files on Zenodo}
The database comes in two zip archives, ``opxrd.zip'' and ``opxrd\_in\_situ.zip''. The latter contains the in-situ data with highly correlated patterns recorded through time series measurements. Within the .zip archives patterns are saved as .json files grouped in folders indicating the contributing institution. If an institution contributed data from several projects, the contributed data is further divided into folders indicating the research project. These research project folders are labeled alphabetically in the order they are introduced in Section~$3$. 
Each .json file contains a pattern recorded from an X-ray diffraction experiment. If available, the composition and structure of the investigated sample and experiment conditions are also included in this file. Patterns belonging to time series measurements are labeled with filenames that indicate the measurement series they belong to and their order in that series. 

\section{opXRD Python library usage}

The opXRD Python library allows the dataset to be accessed through one simple command: \pyth{OpXRD.load(root_dirpath)}. If the database is locally available under \pyth{root_dirpath} this command loads the library from this location. If the database is not available locally at this location, the database is automatically downloaded to \pyth{root_dirpath}. 

\pagebreak

\section{Combined pattern plots}
\begin{figure}[!htb]
    \centering
    \includegraphics[width=0.95\linewidth]{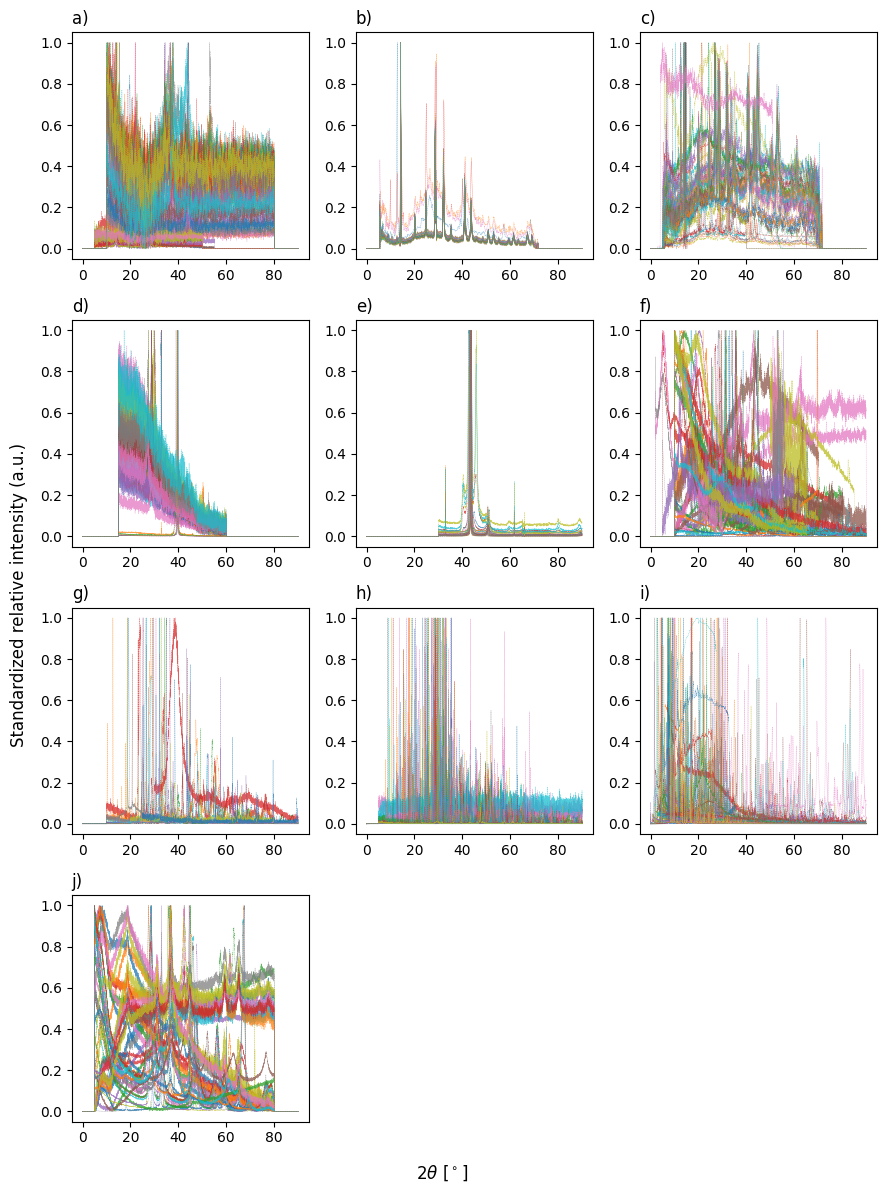}
    \caption{50 randomly chosen X-ray diffraction patterns from each contributed dataset. The figure shows data from the following datasets: a) EMPA, b) LBNL-A, c) LBNL-B, d) LBNL-C, e) USC, f) INT, g) HKUST-A, h) HKUST-B, i) CNRS, j) IKFT.}
    \label{fig:combined}
\end{figure}

Figure~$\eqref{fig:combined}$ shows 50 randomly selected samples of the X-ray diffraction patterns found in each of the research projects contributed to the opXRD database.

